\documentclass{emulateapj}

\lefthead{Ben-Jaffel}
\righthead{HD209458b: spectral, Spatial, and Time properties}

\def \lya {Ly-$\alpha$} 
\def \sn {{\it S/N}}
\def \bjl  {BJ07}
\def \vma {VM03}
\def \vml {VM08}

\begin{document}

\title{Spectral, Spatial, and Time properties of the hydrogen nebula around exoplanet HD209458b}


\author{Lotfi Ben-Jaffel\altaffilmark{1}}
\affil{Institut Astrophysique de Paris, Universit\'e Pierre \& Marie Curie, CNRS (UMR 7095), 98 bis Blvd Arago,
 75014 Paris, France; bjaffel@iap.fr}
\altaffiltext{1}{Visiting Scholar, Department of Applied Science, UC Davis, CA 95616}

\begin{abstract}

All far ultraviolet observations of HD209458 tend to support a scenario in which the inflated hydrogen atmosphere of its planetary companion strongly absorbs the stellar \lya\ flux during transit. However, it was not clear how the transit absorption depends on the selected wavelength range in the stellar line profile, nor how the atomic hydrogen cloud was distributed spatially around HD209458b. Here we report a sensitivity study of observed time and spectral variations of the stellar flux. In particular, the sensitivity of the absorption depth during transit to the assumed spectral range in the stellar line profile is shown to be very weak, leading to a transit depth in the range $(8.4-8.9)\%\pm 2.0\%$ for all possible wavelength ranges, and thereby confirming our initially-reported absorption rate. Taking the ratio of the line profile during transit to the unperturbed line profile, we also show that the spectral signature of the absorption by the exoplanetary hydrogen nebula is symmetric and typical of a Lorentzian, optically thick medium. Our results question the adequacy of models that require a huge absorption and/or a strong asymmetry between the blue and red side of the absorption line during transit as no such features could be detected in the HST FUV absorption profile. Finally, we show that standard atmospheric models of HD209458b provide a good fit to the observed absorption profile during transit. Other hybrid models assuming a standard model with a thin layer of superthermal hydrogen on top remain possible. 

\end{abstract}


\section{Introduction}

Just like our Sun, the \lya\ line profile of HD209458 shows a double-horn emission with the distinguishing feature of intervening interstellar gas strongly absorbing the line core. From the signal left at the Earth's orbit, \cite{vid03} (hereafter \vma ) initially claimed that HST/STIS medium resolution observations of HD209458 are showing a strong absorption of 15\%  during transit and that a large asymmetry appears on the blue side of the \lya\ absorption profile. Those authors used the 15\% drop-off of the stellar signal during transit as an indication of the presence of a hydrogen cloud that extends beyond the Roche lobe of the planet. The next step for \vma\ was to use the asymmetry claimed in the blue side of the line during transit as an indication that the evaporated gas is accelerated by the radiation pressure from the star, an acceleration that results into the build-up of a cometary shape of the nebulosity that escapes from the planet away from the star. \\

Later on, \cite{hol08} rejected the acceleration process by radiation pressure, arguing that neutrals will not have the time to be accelerated by the radiation pressure before they get ionized. Furthermore, they proposed that observations thus far reported by \vma\ on HD209458b could be explained by the presence of a population of energetic neutral atoms produced by charge exchange between the neutral atmosphere and the stellar wind ions beyond an obstacle region corresponding to the planetary magnetopause assumed around $\sim 4.2 $ R$_p$ from the center of the planet. In the meantime, we revisited the HST/STIS medium resolution archive data to report that exoplanet HD209458b absorbs $\sim 8.9\% \pm 2.1\%$ of the parent star \lya\ flux during its transit, much less than previously reported by \vma\ (\cite{ben07a}, hereafter {\bjl}).  We also reported that no sign of asymmetry could be found in the stellar line profile during transit, a finding that directly contradicts the claims initially reported in {\vma}. Soon thereafter, the absorption drop-off reported by {\bjl} for a specific wavelength range was used out of the spectral information provided in Fig.3 of \bjl\ to claim that evaporation was strengthened with a factor $\sim 2$ asymmetry in the absorption between the blue ($9.8\%\pm 1.8\%$) and red ($\sim 5.2\%\pm 1.0\%$) sides of the \lya\ line during transit (\cite{vid08} hereafter {\vml}). In any case, it appears that our initial selection of the wavelength window to estimate the transit absorption created some confusion as to the reading of the data analysis presented in {\bjl}, and this confusion adversely affects efforts to discriminate between the processes thus far proposed to explain the hydrogen distribution around HD209458b ({\vma}, \cite{mun07,sch07,hol08}). In the following, we revisit available \hbox{\lya\ } HST/STIS medium resolution archive data to conduct a sensitivity study that should dissipate the existing confusion, an important step in formulating clear observational constraints on the steady-state properties of the hydrogen nebula around HD209458b. \\

\begin{table*}
\begin{center}
\scalebox{1.}{
\begin{tabular}{lcccccc}

Dataset name &	Program ID & Date Obs.& Time Obs. & Start time - TCT (s) & Duration (s) & Earth's Velocity \\
O4ZEA4010 & 7508 & 2001-08-28 &	06:50:07 & -7980.28  & 2600.	& 2.05 \\
O4ZEA4020 & 7508 & 2001-08-28 &	08:26:25& -2202.29  &	2600  & 2.05\\
O6E201010 & 9064 & 2001-09-07 &20:35:27	& -8075.11	& 1780.	& -2.59\\
O6E201020 & 9064 & 2001-09-07 &22:12:36	& -2246.12	& 2100.	&  -2.62\\
O6E201030 & 9064 & 2001-09-07 &23:48:54	& 3531.89	  & 2100. & -2.65\\
O6E202010 & 9064 & 2001-09-14 &	21:11:51& -10167.67	& 1780.	& -5.68\\
O6E202020 & 9064 & 2001-09-14 & 22:42:08 & -4750.68	& 2100.	&  -5.71\\
O6E202030 & 9064 & 2001-09-15 &00:18:24	& 1025.31	  & 2100. &   -5.74\\
O6E203010$^*$ & 9064 & 2001-10-20 & 02:50:03	& -11258.45	& 1780.	&  -19.30\\
O6E203020$^*$ & 9064 &2001-10-20 & 04:19:45	& -5876.47	& 2100.	& -19.32\\
O6E203030$^*$ & 9064 & 2001-10-20 & 05:55:59	& -102.45	  & 2100.	& -19.34\\

\end{tabular} }
\end{center}
\caption{HST/STIS data set on HD209458 used in this study. All observations were obtained with the G140M grating and the 52"x0.1" long slit. Transit central time (TCT) is defined by 2,452,826.628521 HJD (\cite{knu07}). Earth's velocity is projected along line of sight in the heliocentric reference frame. ($^*$) indicate data set that shows most time variability. \label{tbl1}} 
\end{table*}

Data analysis is carefully described in section 2 so that light curves may be extracted (in selected wavelength windows within the stellar line) and studied in section 3.  In section 4, time variability of the HD209458 \lya\ emission is examined and further evidence is provided on the variable behaviour of the HD209458 emission line, a property that may not be well-described by proxies derived for the solar \lya\ line. The absorption spectral profile is then derived as the ratio between the stellar flux during transit to the unperturbed spectrum of the star with no transit effect.  Finally, the adequacy of models thus far proposed to explain the hydrogen distribution around hot giant exoplanets is discussed in the light of the constraints derived here, with new solutions proposed for the H nebula distribution.

\section{Observations and data analysis}

Four HST visits during the planetary transit of HD209458 are considered, each composed of two to three long exposures ($\sim 2000\, {\rm s}$) distributed in time over the transit period (e.g., Table 1). All exposures were obtained with the HST/STIS G140M medium resolution grating using the 52"x0.1" slit in the time-tag mode. It is important to emphasize that the time-tag mode is a powerful tool that keeps track every $125\times 10^{-6}\, {\rm s}$ of photon events during each exposure. This mode is generally used to improve the spectral and imaging resolutions by correcting for the different motions that may affect the recorded signal (\cite{bro02}). 
Doppler shifts due to HST motion, thermal drifts, and spacecraft jitter/drift are the main sources of motions. In the standard mode (ACCUM) used in several studies ({\vma}, {\vml}), correction is made only for the Doppler motion. For the other sources of motion, we used the time-tag mode, which offers the possibility of cancelling  these motions by taking short sub-exposures that may be cross-correlated. In addition, with its high time resolution, the time-tag mode also offers the opportunity to monitor the time variation of the recorded signal, allowing the study of fast events and/or accurate correction for variable features by also considering short time exoposures. \\

\begin{figure}
\includegraphics*[scale=0.6]{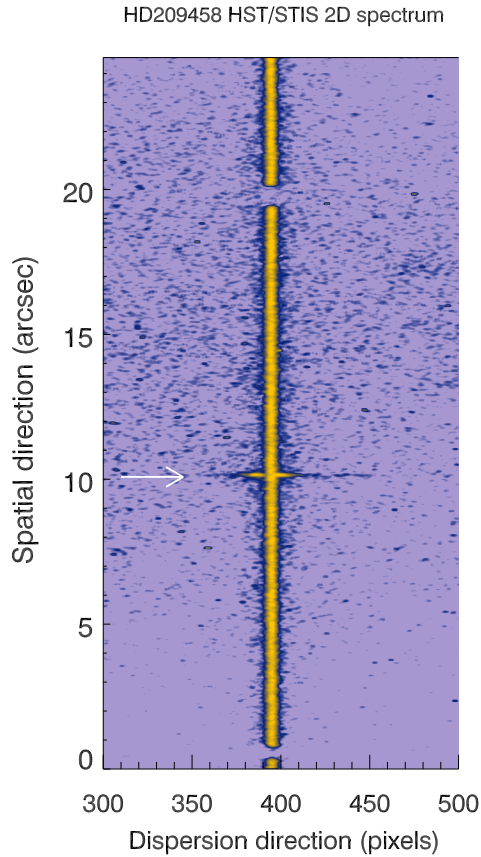}
\caption{ A sample 2D \lya\ spectro-image of HD209458 as obtained by HST/STISG140M medium resolution grating using the long 52"x0.1" slit. The STIS MAMA detector has a $\sim 25"\times 25"$ size, making each pixel covering 0.024" over the sky. The dispersion direction is horizontal so each line of the image is a spectrum.  The spatial direction is vertical. The position of the stellar spectrum along the slit is indicated by an arrow. The bright vertical band is the sky background distribution over the sky. Two gaps, corresponding to the STIS fiducial bars (\cite{bro02}), appear at the bottom and top of the slit. \label{fig1}}
\end{figure}

Following {\bjl}, our set of eleven exposures are sampled here using a time bin of $300\, {\rm s}$ from the available time-tag information. Next, each sub-exposure is calibrated through the STIS pipeline. Because the HST/STIS/G140M grating is a spectro-imaging instrument, every exposure provides an image of the sky along the slit and the corresponding spectra on the cross dispersion direction (Fig. 1). Compared to high resolution observations (\cite{woo05}), the overlap between the stellar and sky signals is worse in the medium resolution, thereby making the corresponding correction more difficult (Fig. 2a). On average, the altitude of the HST orbit is high ($\sim500$ Km) but still within the Earth's hydrogen geocorona. Therefore, any observation is contaminated by the emission and absorption from that upper layers of the Earth's atmosphere and by the absorption and emission from the interstellar medium. Insofar as the star is a point-like source that has few pixels extented over the slit spatial direction (Fig. 1), the other areas of the slit have only the sky background emission which may a priori be used to correct for that contamination. As shown in Figure 2b, the difficulty of this correction comes from the uncertainty about whether the geocorona's signal strength when estimated from a detector sector  along the STIS slit is different from the one where the stellar signal was recorded. Comparing the sky background signal from different sectors of the detector along the slit and for different conditions of observation, we confirm the conclusion of {\bjl} that the STIS MAMMA detector has an inherent non-uniformity corresponding to an incompressible uncertainty of $5$\% on extended sources that is comparable to photon statistical errors.\\

 At this level, the correction for the sky contamination is not yet complete. Indeed, we still need to correct for the absorption by the geocoronal gas. While it may not be necessary when the target's signal is nearly zero in the wavelength range around the spectral position of the \lya\ line in the reference frame tied to Earth  even so, we must check how the line spread function (LSF) of the instrument expands the absorption signature over the spectral domain of interest.  In addition, the spectral position of the geocoronal absorption changes from one exposure to another in the heliocentric reference frame depending on the day of observation (see Table. 1). Assuming standard models for the Earth's upper atmosphere, the \lya\ geocoronal opacity  was evaluated for every exposure before the corresponding absorption profile is derived, convolved with the STIS/G140M LSF, and then applied to the signal after the sky background emission was subtracted (\cite{ben07b,cha87}). Fortunately, most of the geocoronal absorption takes place within the spectral domain that we excluded from our data analysis so that its impact is less than 1\% in the useful domain. Finally, because the interstellar absorption does  not really change from one exposure to another, and should not affect our analysis of the transit effect, no correction is applied for this feature..\\
 
After all corrections were applied, the whole data set resulted in 53 bins time series of the HD208458 planet-star system versus the orbital phase angle that is quite similar to our initial time series within the range of statistical errors (e.g., Fig.~\ref{fig1} of {\bjl}). To ensure that all the effects cumulated (LSF, long slit, thermal motion, etc.) did not corrupt the stellar signal per wavelength pixel at the $1$\% level, the wavelength domain of contamination by the sky background was defined as $[121.541, 121.584]$ nm, the same as in {\bjl}. 


\section{Light curves sensitivity to wavelengths domain}
To obtain the trend of the planetary transit in \hbox{\lya\ }, {\bjl}  integrated the time series spectra in the range $[121.483, 121.536]$ nm on the blue side of the stellar line together with the range $[121.589, 121.643]$ nm on the red side. Accumulating the signal inside the planetary transit period, {\bjl} previously derived a $\sim 8.9\%\pm 2.1$\% drop-off of the stellar \lya\ intensity. The question then is: does the absorption rate depend on the selected wavelength range? For example, using low resolution HST/STIS observations of HD209458, \cite{vid04} found an absorption rate of $5.2\% \pm 2$ \% for neutral hydrogen, a rate much smaller than they derived when using HST/STIS medium resolution observation. Recently, {\vml} claimed that the difference between our estimate ($8.9\%\pm 2.1\%$) and their initial absorption rate ($15\%\pm 4\%$) comes from the distinct integrated wavelength ranges considered. These authors also claimed that they can reconcile medium- and low-resolution observations  using the same argument of difference in the wavelength range namely, full line in the low-resolution case compared to restricted windows around the stellar line peaks in the medium-resolution observations analysis. Therefore, it is important to verify these claims and to estimate the sensitivity of the  transit absorption to the selected wavelength range.\\

\begin{figure}
\epsscale{1.25}
\includegraphics*[scale=0.45,trim= 0 0 20 0]{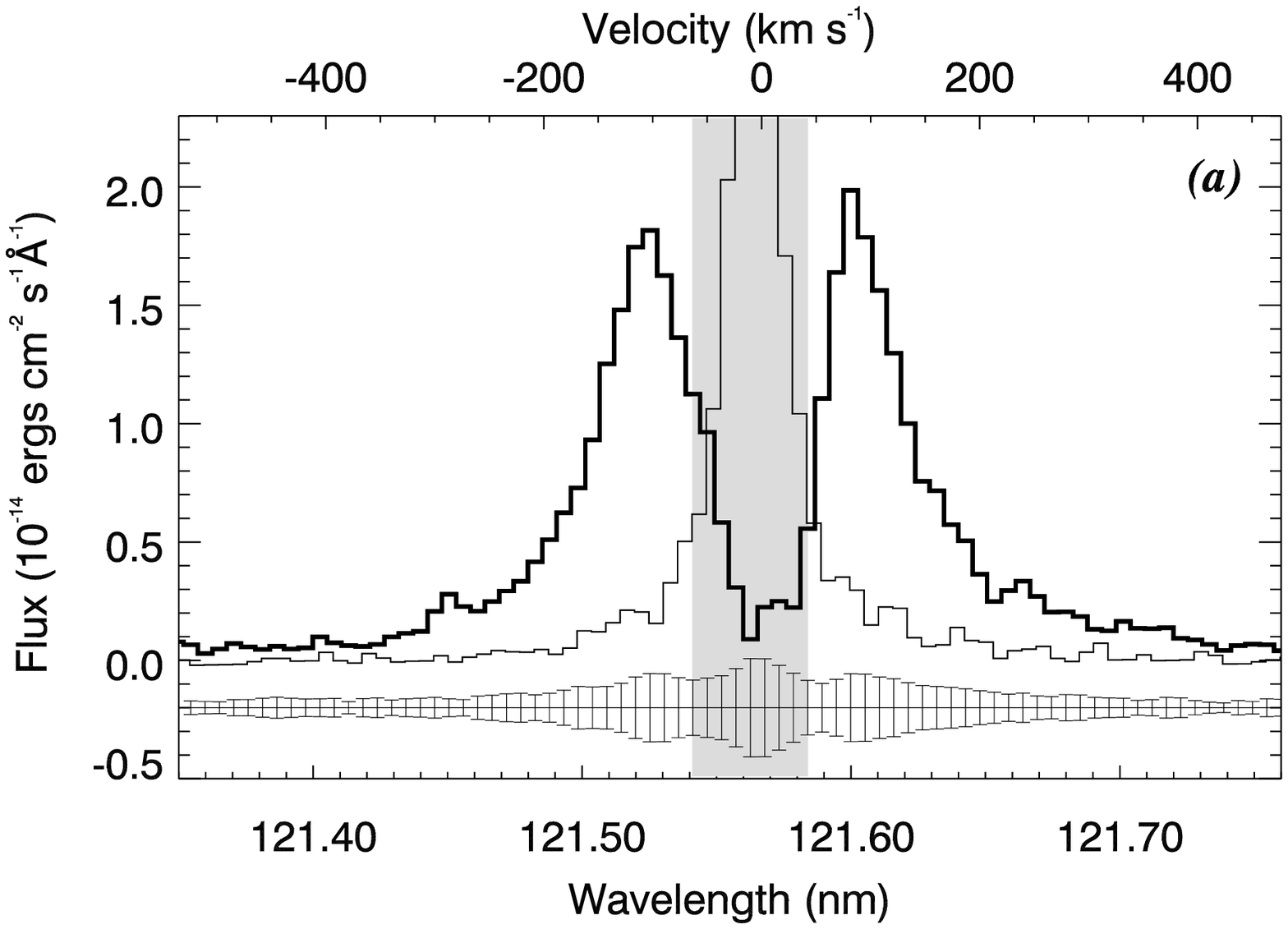}
\epsscale{1.4}
\noindent 
\includegraphics*[scale=0.52,trim= 50 0 0 0 0]{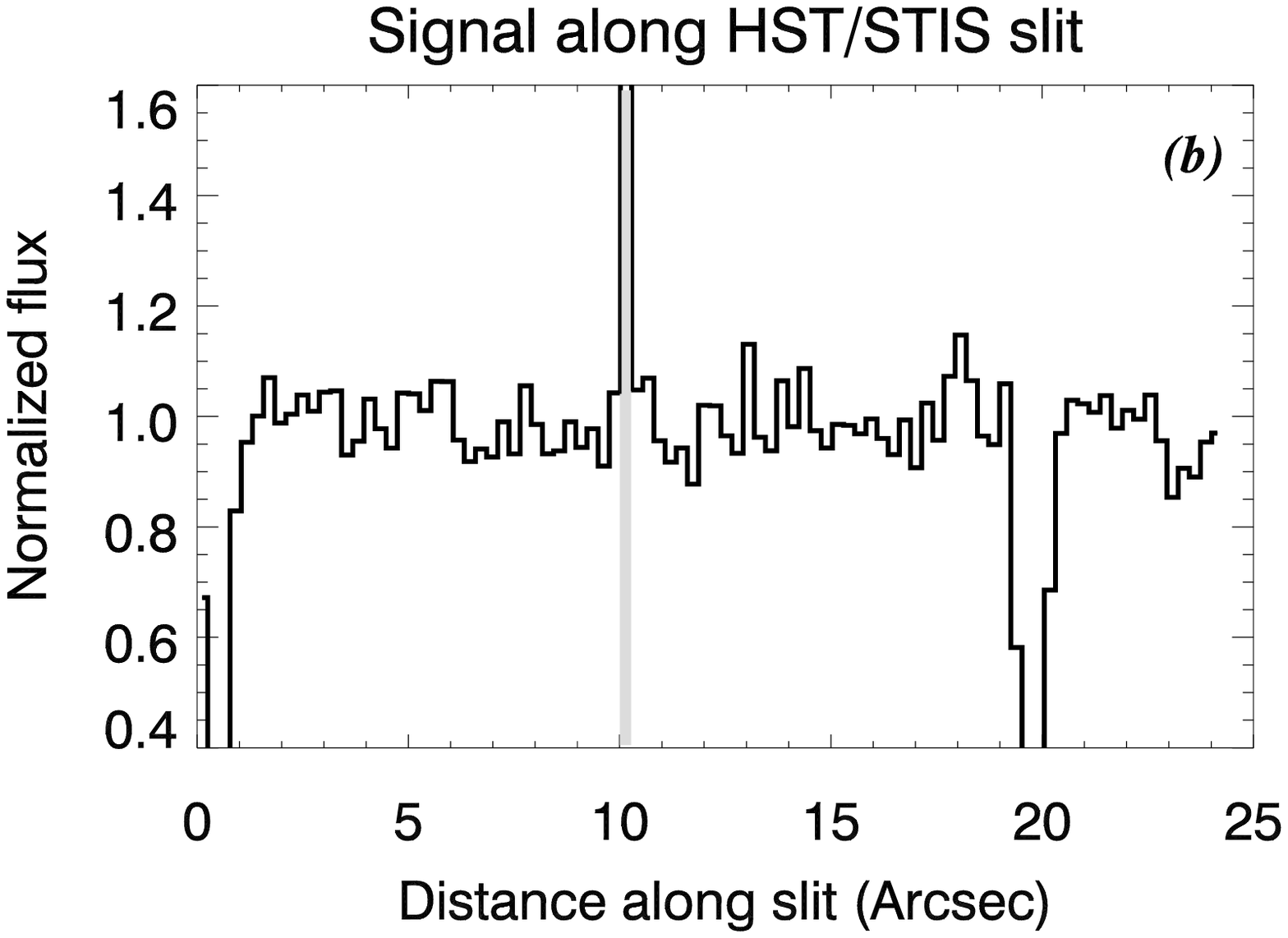}
\caption{{\bf (\it \bf a)} A sample spectrum of HD209458b \lya\ emission line observed by HST/STIS/G140M medium resolution after correction for all sky background contaminations (thick line). The sky background contaminations (including the geocoronal and the interplanetary \lya\ emissions for the date of observation; see Table. 1) is displayed (thin line) to highligh its line wings contamination due to the effect of the line spread function of the instrument. The spectral domain of the sky background contamination is shown as a shaded area. {\bf (\it \bf b)} Sky and stellar integrated flux versus distance (arcsec) displayed along the HST/STIS/G140M slit.  The flux was evaluated by integration over the spectral domain ($[121.483,121.643]$) englobing the spectral extent of the sky emission and part of the stellar line. The signal spike indicated by a shaded area around positions $\sim 10$" corresponds to the stellar line. The signal has been binned to eleven pixels corresponding to the default spatial extent of a point-like source. The signal gaps respectively around positions $\sim 0-1$" and $\sim 20$" correspond to the location of the fiducial occulting bars of STIS. The variation (5\%) of the signal along the slit that results from the non-uniformity of the STIS MAMA detector is comparable to the statistical noise level outside the spectral domain $[121.541, 121.584]$ nm of contamination by the sky background signal.  \label{fig2} }
\end{figure}

Starting from the time series of the \lya\ line emission  (e.g., Figure~1 in {\bjl}), we have generated light curves (LCs) for several wavelength windows, including those considered by {\vma} and more recently by {\vml}. 
As an initial step, we tried to compare LCs corresponding separately to the blue and red line sides (Fig.~3). Accumulating the signal from three bins inside transit with the corresponding time weights, our analysis shows an absorption $\sim 8.5\%\pm 2.5\%$ for the blue side and $\sim 8.2\%\pm 2.5\%$ for the red side (see Table 2 for details). As shown in Figure 3, the two LCs behave similarly over time, except for the last bin around time $\sim 5000\, {\rm s}$ where the red side seems even more absorbed. Despite the noise level, the observed trend for the two LCs supports the idea that both sides of the stellar line are absorbed almost equally during transit. By contrast, for the same wavelength domains, {\vml} recently reported absorptions of $9.8\%\pm 1.8\%$ and $\sim 5.2\%\pm 1.0\%$, respectively, for the blue and red sides of the line, suggesting a rather large asymmetry between the two sides. Evidently, our results differ significantly from those of {\vml} when considering the two sides of the stellar line separately. However, when merging signals from the two sides of the line, the corresponding absorptions seem comparable for the two analyses, yet this similarity ($8.4\%\pm 1.8\%$ here and in {\bjl} compared to $7.3\%\pm 2.\%$ in {\vml} ) is only incidental. \\

\begin{figure}
\epsscale{1.25}
\includegraphics*[scale=0.44,trim= 40 0 0 0]{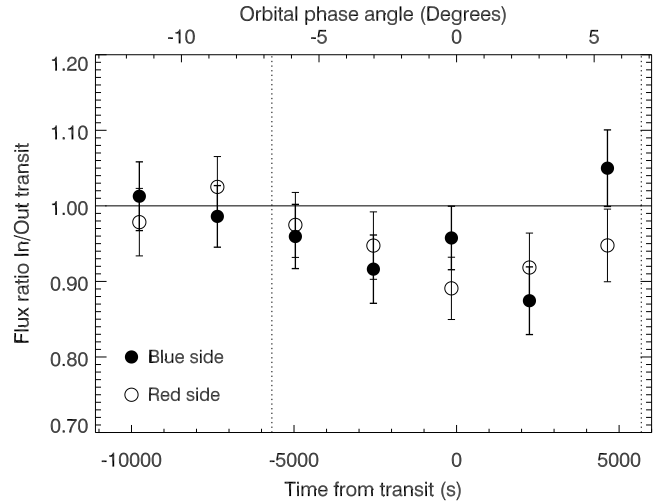}
\caption{Light curves obtained from our time series (as shown in Fig. 1 of {\bjl}) by integration of the signal in the spectral windows [121.483, 121.536] nm for the blue side (filled circle) of the stellar line and [121.589, 121.643] nm for the red side (open circle). Despite the noise level, the two curves behave similarly during transit with a total obscuration of $\sim 8.5\%\pm 2.5\% $ for the blue side and $ \sim 8.2\%\pm 2.5\%$ for the red side (see text). If any difference, it would be in the opposite direction of {\vml}'s claim, as indicated by the time bin around $\sim 5000 {\rm s}$ where the red side seems more absorbed than the blue one. For reference, the beginning and end of the visible transit are shown by vertical dotted lines. \label{fig3} }
\end{figure}

In a second step, we varied the wavelength domain of integration over the stellar line. As shown in Figure 4, both our selected wavelength window and the one of {\vma} give quite similar LCs with a transit absorption rate around $\sim 8.4\% \pm 1.8\%$ for our spectral domain and $\sim 8.85\% \pm 2.2$\% (compared to $15\% \pm 4\%$ in \vma\ ) for the domain selected by {\vma}. In addition, when considering the whole stellar line, our absorption rate remains around $\sim 8.9\%\pm1.8\%$, consistent with our initial estimate and also with recent analysis using HST/ACS observations (\cite{ehr08}). \\

\begin{figure}
\epsscale{1.25}
\includegraphics*[scale=0.44,trim= 40 0 0 0]{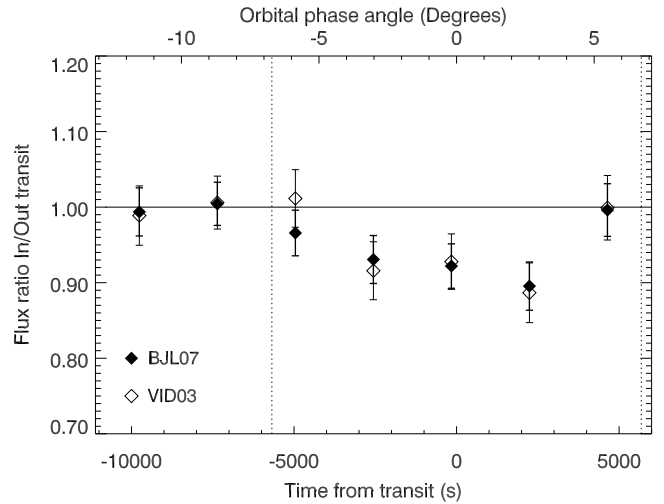}
\caption{Light curves obtained from our time series (as shown in Fig. 1 of {\bjl}) by integration of the signal in the spectral windows [121.483, 121.643] nm (filled diamond, {\bjl}) compared to the one obtained from the wavelength windows [121.515, 121.61] nm (open diamond, {\vma}). A total obscuration of $\sim 8.4\%\pm 1.8$\% is derived for the first domain and $8.8\%\pm 2.2\% $ for the second one during the planetary transit. The closeness of the two absorption rates rules out any large spectral dependency as claimed by {\vma} and {\vml}. For reference, the beginning and end of the visible transit are shown by vertical dotted lines. \label{fig4} }
\end{figure}

What, then, is the physical origin of this weak sensitivity of the absorption during transit to the selected wavelength range? Before we proceed to answer this question, in the following we study the time variability of the stellar in order to explore its potential impact on the transit absorption line profile.

\section{Time variability of the HD209458 \lya\ signal}

Time variability of the stellar signal is considered here as a potential cause for the discrepancy between the different data analysis related to the HST/STIS/G140M medium resolution observations. \vma\ and \vml\ used the solar \lya\  variability as a proxy to catch the variability of the \lya\ emission line of HD209458 (solar type star). This approach may not be adequate because no two stars of the same type behave similarly over short time scales. Moreover, HD209458 has a nearby orbiting exoplanet that may induce tidal and electromagnetic modifications of the stellar signal, effects that do not exist for our sun. As a matter of fact, HD209458 was previously suspected to have a relatively moderate chromospheric activity from CaII H and K lines that were recorded over full orbits of the system (\cite{shk05}). More recently, time variability of the stellar \lya\ flux was noted and showed both for a specific wavelength range and at different orbital phases (see Fig. 2 \& 3b in {\bjl}). However, the noise level was relatively high making difficult the diagnostic on the origin of the signal variability. Another possible way to characterize time variability is to carefully inspect the individual data set so far obtained by comparing the stellar signal at different phase positions to test any correlation with the orbital position of the planet. Such correlation may result from unkown electromagnetic interactions suspected to operate between the planet and the star during transit (\cite{cun00,cra07}), much like what we know from the Jupiter-Io or Jupiter-Ganymede systems (\cite{cla02}).\\

\begin{figure}
\epsscale{1.25}
\includegraphics*[scale=0.44,trim= 40 0 0 0]{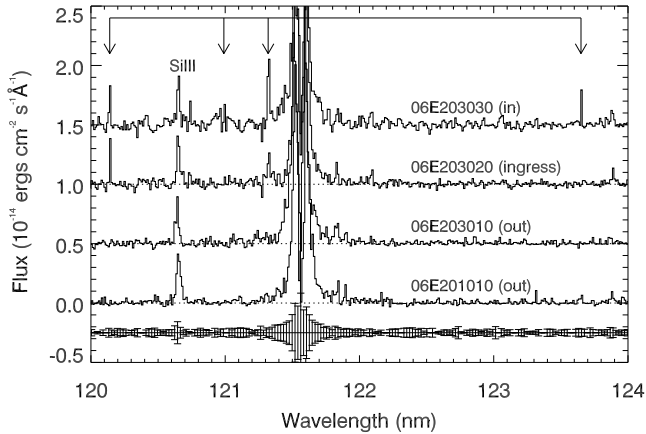}
\caption{HD209458 spectra obtained from data files (O6E203010, O6E203020, O6E203030) for different orbital positions of the planetary companion. A "quiet" spectrum from the remaining HST data set (eg Table 1) is shown for reference. The spectra are rebinned by 2 at a spectral resolution of $\sim 26 {\rm Km s^{-1}}$. Several spectral features (indicated by arrows) appear in the O6E2030x0 spectra but could not be found in the other exposures. Moreover, the strength of these features changed over time as if related to the orbital position of the exoplanet. However, because the phase positions are set by the HST observing schedule, the transit period is not fully covered, thus making it difficult to correlate the signal variation with the transit effect and the related orbital positions of the exoplanet. \label{fig5} }
\end{figure}

 A careful inspection of the four visits thus far available in the archive reveals that, indeed, the data subset obtained during the third transit of the {\vma\ }observing program (data files O6E203010, O6E203020, and O6E203030; see Table 1) show extra features  that are not observed in the other visits of the HD209458 system (Fig. 5). Some of the observed features are well above the noise level and are suspected to correlate with the planet orbital position, particularly for the spectral feature that appears around $\sim 121.3$ nm in the far blue side of the stellar \lya\ line. If we restrict the analysis to the solely \lya\ line profile in the range $\pm 200$ km/s from the line center, it is difficult to descriminate a time variability from statistical variations at low {\sn\ }. Yet 
we can estimate such time variability by comparing integrated fluxes from the different exposures as shown for example in Fig. 3 of \vma\ and Fig. 2 of \vml . In these figures, the data points corresponding to the third transit (the so called C points) behave strangely with a departure of up to 3 $\sigma $ far from the theoretical transit light curve. 

All these indications tend to support that data set O6E2030X0 bears an unusual variability that is not found in the other exposures. The extra features are not related to any Earth's energetic particles activity along  the HST orbit, as all observational parameters are comparable to the other data sets.  We suspect that the data set O6E2030x0 may have corrupted the diagnostic on the transit depth and absorption profile in \vma\ analysis. In our case, the impact of the same data set was largely diminished by the splitting of the exposures into our time series and by having extra observations (data files O4ZEA40X0, see Table 1) strengthening the weight ($\sim 25$\% more) of the pool of "quiet" exposures.  Finally, the uniqueness of the disturbed data set O6E2030X0 and the noise level make it difficult to definitely assess a correlation with the orbital position of the planet, a question that remains open until future observation of the HD209458 system can be obtained after the FUV capabilities of the HST are restored. 

\section{Absorption line profile during Transit}

In practice, to obtain a self-consistent absorption profile, both steady-state unperturbed (out-of-transit) and in-transit line profiles are required and will be processed by iteration. 
In the present case, the iteration process converged in only a few steps, providing the absorption profile as shown in Figure 6. The obtained absorption spectral profile is typical of damped absorption systems, though the spectral domain around the line center is hidden by, respectively, the interstellar medium absorption and the contamination from the sky background. The first thing to note is that the atmospheric absorption is statistically flat far into the stellar line wings down to $121.5$ nm on the blue and up to $121.63$ nm on the red, including the spectral position of the stellar line peaks. As the peaks' intensity provides the highest contribution to the LC shown in Figures 3-4, it becomes apparent that a flat absorption of $\sim 8.9\% \pm 2.1\%$ proposed in our initial study ({\bjl}) is a rather good approximation. Beyond that frequency domain, it is difficult to accurately assess the spectral variation due to the noise level. \\

Binning the absorption line profile in wavelengths improves the signal quality, yet it does not reveal any asymmetry between the blue and red sides at a reduced spectral resolution of $\sim 26{\rm km\, s^{-1}}$. As previously shown in Figure 3, fluxes integrated, respectively, on the blue and red sides of the emission line behave quite similarly over time, supporting the contention that both wings are absorbed almost symmetrically during transit in the spectral domain $< 200 {\rm km\, s^{-1}}$ from line center. Any difference, if at all, must be weak and/or occur outside the available spectral domain such that current data noises and/or the sky background contamination hide it.\\

\begin{figure}
\epsscale{1.25}
\includegraphics*[scale=0.44,trim= 40 0 0 0]{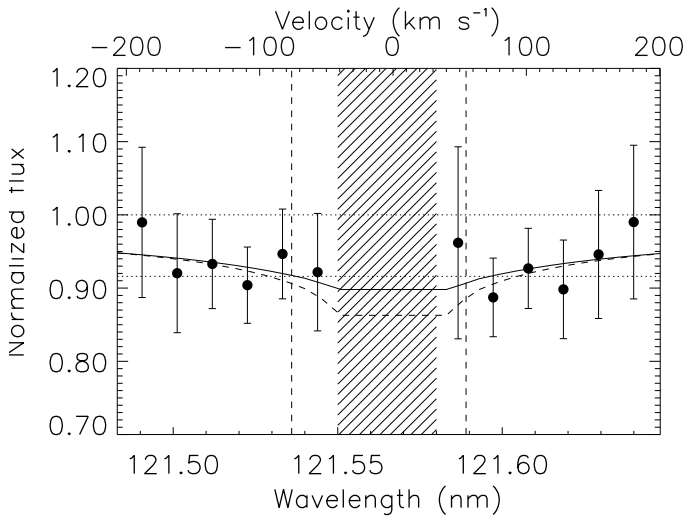}
\caption{Absorption line profile (filled circles) during transit of HD209458b obtained as a ratio between \lya\ line profiles in transit and the unperturbed one. The sky background spectral domain as defined here is indicated by two vertical dashed lines, while the one of {\vma} is indicated by a hashed area. The spectral window was restricted to $\sim \pm 200 {\rm km\, s^{-1}}$ from line center, because the signal becomes much too noisy beyond. A flat absorption rate of $\sim 8.4\% $ fits rather well the obtained absorption profile. Using the standard atmospheric model of \cite{mun07}, a first model fit to the absorption line profile (solid line) has a Lorentzian shape with extended wings. Because data bear statistical and variability noises and the core spectral domain is absent at Earth's orbit, other scenarios of gas distribution may produce an absorption profile (dashed line) that fits rather well with the observed one, and thus cannot be ruled out. \label{fig6} }
\end{figure}

In the following, we take a theoretical approach to verify whether such observational properties of the absorption profile are predictable, particularly its Lorentzian shape (Fig. 6). First, we recall some theoretical background on the opacity properties within a strong conservative line like {\lya }. Typically for a resonance line, the absorption is dependent on wavelengths with strong opacities in the line core and much less in the (absorption) line wings. For a typical temperature of $\sim 10000$K for the HD209458b inflated atmosphere (\cite{mun07,yel04}), the absorption Doppler width is almost $\sim 0.0052$ nm (or $\sim 13 {\rm km\, s^{-1}}$), which is comparable to the spectral resolution ($\sim 13 {\rm km\, s^{-1}}$) of the HST observations used here. Referring back to the HD209458 line profile ({\vma}, {\bjl}), the line's peaks appear at $\sim \pm 100 {\rm km\, s^{-1}}$, or equivalently, $\sim \pm 8$ Doppler units far from the line center, placing them well in the wings of the absorption line profile of the planetary atmosphere. Because of the steep fall off of the flux in the stellar wings, the contribution of the signal around the peaks' positions is dominant in any estimate of the flux attenuation during transit. Also, one should note that for a target that is not spatially resolved, such as transiting exoplanets, the observed signal extinction is a weighted average of absorption from different regions of the exoplanetary extended atmosphere, with the outer layers having the maximal weight. This could be examined by writing the basic definition of the observed (transmitted) stellar line profile during full transit: \\

\begin{equation}
 F(v)={\int\int \exp{(-\tau \Phi(p,\theta,v))} F_{in}(v,\theta).p.dp.d\theta } 
\end{equation}

where $\Phi(p,v))$ is the absorption profile by H atoms versus the velocity $v$ (or wavelength) from the line center, $F_{in}(v)$ is the unperturbed stellar line profile, and $\tau (p,\theta)$ is the opacity along the line of sight that goes through the cloud's volume at the impact parameter $p$ from the star's disc center and a polar angle $\theta$. The integration is evaluated over the whole extent of the stellar disc, allowing various  distributions of the gas around the planet that may lead to the same observed extinction. This degeneracy would not appear if the atmospheric layer were thin compared to the size of the planet.\\

To illustrate the multiplicity of solutions, a standard atmospheric model of HD209458b (\cite{mun07,yel04}) was used. As shown in Figure 6, a rather good fit to the absorption line profile during transit is obtained for the 
{\it DIV1} model of \cite{mun07}. The DIV1 model includes heavy elements contribution to the H production and appears up-to-date regarding chemistry reactions. If using the standard model of \cite{yel04} that does not include heavy elements chemistry, the same good fit is obtained when the H abundance is scaled by a constant factor. Beyond the difference between \cite{mun07} and \cite{yel04} models, the scaling constant may also reflect other processes of H atomic production that are not accounted for in classical photochemistry models. In the past, the study of the thermosphere of giant planets such as Jupiter and Saturn and their satellites showed the need for such enhancement of the H production in order to fit the \lya\ observations thus far obtained by Voyager 1 \& 2, IUE, Galileo, or HST (see, for instance, \cite{ben07b,ben95,cla91,san80}). For example, in the case of Jupiter, a planetwide enhancement ($\sim 3$ times) of the H production is expected from magnetospheric protons and electrons precipitation in the polar regions (mainly H2 or heavy molecules dissociation by energetic particles). For reference, the same process produces the classical auroral emissions. The constituents produced by auroral processes and the input energy and momentum in the polar regions are then transported and distributed over all latitudes/longitudes (\cite{som95,bou05}). As of today, the H abundance and distribution on Jupiter are not yet fully uncovered (\cite{ben07b}). The idea here is that the magnetospheric energy input in the auroral
regions may play a key role in the upper atmosphere of HD209458b's composition and structure. Such effects are not taken into account in atmospheric models thus far published in the literature. On that basis, for
a planet such as HD209458b orbiting at 0.05AU from its parent star, thereby receiving much more energy through its polar regions than expected on Jupiter, the conditions may favor a rather strong enhancement of the H production that may lead to larger atmospheric opacities at \lya\ than shown in Fig. 15 of \cite{mun07}. \\

In the model calculations shown in Figure 6, a Voigt function was assumed for the H atomic absorption profile in order to account for the natural broadening of the \lya\ line. As expected from the line shape, the atmosphere is opaque and consistent with the line-of-sight opacities used (see Figure 15 in \cite{mun07}). The line shape reflects the effect of the broadening in \lya\ line wings by the large H column along the line of sight that leads to a large and relatively flat absorption far from the line center. However, it is important to point out that the interpretation of the absorption level versus wavelengths is not straightforward because the absorption is a weighted contribution from different atmospheric layers located at different distance from the planet's center (e.g., Equation 1). For that reason, we stress that the fit to the absorption profile and the corresponding atmospheric opacities correspond to one possible sketch of the H distribution that may not contain all of the key physical processes at play in the real spatially extended system.  Other scenarios of gas distribution may lead to the same observed absorption profile (e.g. Fig. 6). For example, we are able to find a more extended atmosphere for which the extinction by the exoplanet atmosphere fits rather well with the observed absorption profile (see Fig. 6). As it is impossible to obtain any information on the line core because of the interstellar medium absorption and the sky background contamination, only very high {\sn\ } FUV observation at a level better than $(0.1-1.0)\% $ may disentangle the different processes in play.

\section{Comparison with previous analysis}

At this point, we are able to compare our result with previous analysis that used most of the same data set ({\vma}, {\vml}).
The immediate question is, what could be the origin of the difference in the transit absorption rate as derived for the same wavelength domain
$DW1:$ [v1,v2]=[121.515, 121.61] nm--namely, $8.8\%\pm 2.2\%$ in this study--compared to $15.\%\pm 4\%$ in the case of {\vma} and {\vml}?
{\vml} claimed that a difference in the selected wavelength domains was the cause of the difference in the absorption rates. To support their
claim, those authors derived a transit absorption of $15\%\pm 4\%$ from their analysis for the wavelength domain $DW1$, while a different transit absorption of $7.3\%\pm 2.1\%$ was derived for the wavelength domain $DW2:$ [v3,v4]=[121.484, 121.643] nm. The problem is that the coherency claimed by {\vml} is merely illusory because the provided rates are simply inconsistent. Indeed, one may note that the considered wavelength window
$DW1$ is embedded in $DW2$, with $DW3 :$ [v1,v3] \& [v4,v2]=[121.484, 121.515] nm \& [121.61, 121.643] nm as the difference domain. Considering the out-of-transit stellar line profile as provided by {\vma} and {\vml}, one may wonder: how could an absorption rate possibly be cut by a factor of two when extending the wavelength domain from DW1 to DW2 and only adding the falling-off signal from the very limited DW3 stellar wings? In other words, what could be the transit absorption rate in the complementary domain $DW3$ if 15\% is obtained in the first domain $DW1$ and 7.2\% in the second one $DW2$ as estimated by {\vml}?\\

\begin{table*}
\caption{Sensitivity of the transit absorption rate to the spectral domain} 
\begin{center}
\scalebox{1.}{
\begin{tabular}{lccccc}
Wavelength window (nm); reference &	Published absorption (rate, reference) &	Absorption rate (estimated here) \\
$[121.515, 121.610]^{a}$; ({\vma}) & $ 15\%\pm4\% $ ({\vma})  & $8.8\%\pm 2.2\% $	 \\
$[121.483, 121.643]^{b}$; ({\bjl}) &  $8.9\%\pm 2.1\% $ ({\bjl}); $7.3\%\pm 2.1\% $ ({\vml}) & $8.4\%\pm 1.8\% $	 \\
$[121.483, 121.536]$; ({\bjl}, blue side) & $9.8\%\pm1.8\% $ ({\vml}) & $ 8.5\%\pm 2.5\% $	 \\
$[121.589, 121.643]$; ({\bjl}, red side) & $5.2\%\pm 1.0\% $ ({\vml}) & $8.2\%\pm 2.5\% $  \\
$[121.515, 121.550]$; ({\vma}, blue side) &   ---  & $9.3\%\pm 3.3\% $ \\
$[121.580, 121.610]$; ({\vma}, red side)  &  ---   & $8.3\%\pm 2.9\% $ \\
\end{tabular}}
\end{center}
\tablenotetext{}{(a) {\vma} geocoronal window [121.55,121.58] nm excluded. (b) {\bjl} geocoronal window [121.541, 121.584] nm excluded.}
\end{table*}

Starting from the definition of the transmitted flux in Equation 1, we write that the absorption $A_{DW1}$ integrated over the wavelength domain $DW1$ is a weighted sum of the absorption $A_{DW2}$ integrated in the domain $DW2$ and $A_{DW3}$ in the domain $DW3$. For a sake of clarity we introduce the following notation for the signal attenuation $A_{DW}$  corresponding to a prescribed wavelength domain $DW$ :
\begin{equation}
 A_{DW} = 1- I_{DW} 
\end{equation}
 where 
\begin{equation}
 I_{DW} = {I^{N}_{DW}\over I^D_{DW}}
\end{equation}
 is the fraction of transmitted signal with $$I^N_{DW}=\int_{DW} F(v) dv$$ and
\begin{equation}
 I^D_{DW}=\int_{DW} F_{in}(v) dv
\end{equation}
In above equations, the subscripts $N$ stands for numerator and $D$ for denominator, and $F_{in}$ is the unperturbed stellar flux. Note that the contribution to the integration from the sky background wavelength domain is set to zero.\\
Starting with the most extended $DW2$ domain, we may write: 
\begin{equation}
 I_{DW2} = {\int^{v4}_{v3} F(v) dv\over \int^{v4}_{v3} F_{in}(v) dv} \end{equation}
Splitting the integration in the numerator over the different domains, we obtain
\begin{equation}
I_{DW2} = {(\int^{v1}_{v3}+ \int^{v2}_{v1}+\int^{v4}_{v2})F(v) dv\over \int^{v4}_{v3} F_{in}(v) dv} \end{equation}
Taking into account the definition of the rates in the $DW2$ and $DW3$ domains, we can write:
\begin{equation} I_{DW2} = {I_{DW1}\times I^D_{DW1} +  I_{DW3}\times I^D_{DW3} \over I^D_{DW2} } \end{equation}
We now may derive the absorption in the $DW3$ wavelength domain :
\begin{equation} I_{DW3} = { I_{DW2}\times I^D_{DW2} - I_{DW1}\times I^D_{DW1} \over I^D_{DW3}} \end{equation}

All integrals $I^D_{DW}$ are known and directly derived from the unperturbed (out-of-transit) line profile.  Now, 
using the numbers provided by {\vma } and {\vml } for the attenuation in the different wavelength domains of interest here, namely that $A_{DW1} = 15\% \pm 4\%$ (or $I_{DW1} = 85\% \pm 4\%$) and $A_{DW2} = 7.3\%$ (or $I_{DW1} = 92.7\% \pm 2\%$), we can derive the fraction of the signal transmitted in the $DW3$ wavelength domain should be $I_{DW3} \sim 108\% $. This means that in order to 
make consistent the 15\% and 7.2\% rates derived by \vml\ respectively for the $DW1$ and $DW2$ spectral domains, an enhancement of 8\% of the stellar signal is required during transit in the wavelength window $DW3$.
Surprisingly, when the absorption rate proposed by \vma\ is divided by a factor 1.08 (8\% enhancement), then we recover an absorption rate of 8.7\% for the $DW1$ spectral domain, consistent with our present estimation (see Table~2). We believe that the normalization used by \vma\  to compensate stellar variability by using the stellar wings signal as a proxy may be one of the causes of the discrepancy between their results and ours. Indeed, \vma\ and \vml\ repeatedly claimed that because most of the absorption should, in their estimation, occur in the line core, then the stellar \lya\ line wings (out-domain in \vma\ definitions) should not show any spectral absorption. As shown here (e.g., Fig. 6), the transit absorption profile has very extended wings with a substantial absorption up to the out-domain used by {\vma}, which casts doubt upon their approach to estimating the transit attenuation and to handling stellar time variability using the line wings' signal as a reference.

\section{Discussion}
Despite serious limitations of current technology to resolve exoplanets from their parent stars, we learned a lot on their atmospheric structure thanks to the transit technique. For the extrasolar planet HD209458b, we are close to sketching its atmospheric structure over several scales with detection, respectively, of neutral sodium in the lower atmosphere (\cite{cha02}), very extended corona of hydrogen in the upper atmosphere, and an intermediate layer of temperature jump (\cite{bal06}). For exoplanets orbiting very close to their host stars, these results confirm the scenario of atmospheric inflation by the huge stellar flux, a fact that raises the question of atmospheric evaporation and the survival of these planets over several stages of stellar evolution. \\

Therefore, it may be worthwhile to discuss the adequacy of thus-far proposed models to explain the hydrogen distribution around hot giant exoplanets in the light of the constraints derived here and to propose new hints that may help investigate this fundamental problem. Since \vma\ disseminated their results on HD209458, many studies followed regarding the mass loss from exoplanets, particularly that class of planets that closely orbit to their parent stars. Initially, \vma\  proposed evaporation followed by radiation pressure acceleration of hydrogen atoms beyond the Roche lobe leading to a comet-like shape of the hydrogen nebula that escape away from the star. As we know from \bjl\ and the present study, this scenario is not confirmed by the HST/STIS \lya\ observations because no signature of asymmetry could be found as initially claimed by {\vma}. The second difficulty of the \vma\ model was pointed out by \cite{hol08}, namely, that the radiation pressure has not enough time to operate before the atoms are ionized. \\

These conclusions do not mean that evaporation is not operating in the HD209458 system. In fact, the only process being dismissed is the acceleration by radiation pressure because, consistently with the conclusions of the present analysis, it does not seem to be efficient enough to produce the necessary population that must leave a clear spectral asymmetry in the \lya\ absorption profile. \cite{sch07} tried 3D hydrodynamical simulation of the hydrogen nebula around HD209458 and predicted several light curves for the planetary transit, depending on the escape efficiency of the atmosphere. Unfortunately, these authors underestimated the absorption by the hydrogen cloud by assuming a Gaussian absorption profile for the H atoms at {\lya\ }, an assumption that is not adequate for this atomic transition. In all cases, the natural width of the upper level of the H atom induces a natural broadening, described by a Lorentz profile that must be included in the absorption calculations (\cite{mih78,ben88}). When included, this natural broadening should significantly enhance the absorption in the \lya\ line wings and consequently the absorption for the total line. If the velocity distribution of atoms is Maxwelian, then the attenuation is enhanced by the ratio $\sim {x^{-2}\over \exp{-x^2}}$ of a Voigt to a Gaussian function that may be several order of magnitude beyond $x\sim 3$ Doppler units from the line center (\cite{neu90}).\\

In a later study, \cite{hol08} proposed that the inflated atmosphere of HD209458b should be extended beyond the magnetopause region of the planet so that charge exchange with the stellar wind protons would necessarily produce energetic neutral atoms (ENAs). In this process, the size of magnetospheric cavity (obstacle region in \cite{hol08}) is one of the key parameters that defines the ENAs content. The more the obstacle region is extended, the less ENAs are produced. Posing the magnetopause position at $\sim 4.2 {\rm R_p}$, these authors were able to account for the HD209458b transit's absorption features as claimed by \vma\ assuming a stellar wind with a  density of $2\times 10^{3} {\rm cm^{-3}}$ and a bulk velocity $\sim 50 {\ km s^{-1}}$. As far as the conclusions of \vma\ and \vml\ have been invalided, the fit proposed in \cite{hol08} is no more adequate and one may question if other parameters of the stellar wind could account for the constraints derived in \bjl\ and in the present study.  \\

Before we conclude on this question, it is important to discuss some difficulties of the ENAs production model as proposed by \cite{hol08}. Because of the nature of the charge exchange reaction, we must have two populations: one is the ENAs and the second is a family of ions flowing outward in the same direction as the planetary atoms that produced them. In their publication, \cite{hol08} just removed that population from their simulation box. Also, the stellar wind should slow down when approaching the assumed obstacle. A self-consistent formulation of the problem should allow for a cascade of ENAs and ions productions through the charge exchange process (see \cite{ost92}). Such ions will induce their own ram pressure against the stellar wind, pushing away the magnetospheric cavity (the obstacle) outward at larger distances (\cite{rat02}). How great that distance is depends on self-consistent simulations that need to be done. According to \cite{hol08} (Fig. 7 in their supplementary draft), moving the obstacle (magnetopause) further out will reduce the ENAs production and, consequently, the attenuation in the red and blue part of the spectrum. The conclusion here is that the current version of the the ENAs model as proposed by \cite{hol08} lack the coherent treatment of magnetospheric structure and response that is much needed to evaluate whether this process is important in shaping the hydrogen distribution around HD209458b. \\

\begin{figure}
\epsscale{1.25}
\includegraphics*[scale=0.505,trim= 30 0 0 0]{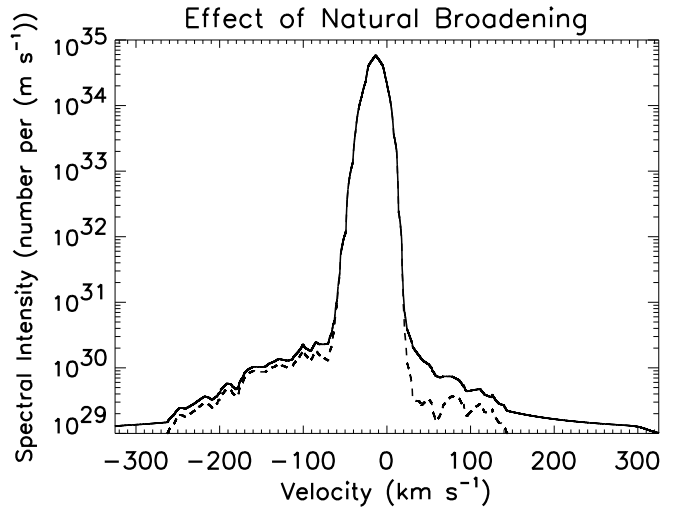}
\caption{(Dashed line) Velocity field of neutral hydrogen proposed by \cite{hol08} as a best fit to account for ENAs production through charge exchange between HD209458b atmospheric hydrogen and impinging stellar wind protons. (Solid line) The same distribution when accounting for natural broadening due to the finite life time of the H second atomic level. The net effect of natural broadening is to spread the initial distribution over a wider velocity range and making its red side stronger ($\sim 5 $ times) and almost symmetric compared to its blue side in the reference frame of the star. This result is independent of the intensity of the distribution that should be normalized to unity to derive the local atomic absorption profile. \label{fig7} }
\end{figure}

Another difficulty the \cite{hol08} model faces is the formulation of the attenuation process of the \lya\ emission due the atomic hydrogen. As in \cite{sch07}, these authors evaluated their attenuation profile during transit while overlooking the fact that the second level of a hydrogen atom has a finite life time that induces a natural broadening of the \lya\ line. The natural broadening is represented by a Lorentz function that describes the absorption profile of H atoms in their rest frame. Therefore, for a gas with a given velocity distribution (like the one shown in Fig. 2 of \cite{hol08}), the absorption profile that should be used for attenuation calculations must be obtained as the convolution between the Lorentz profile and the velocity distribution (\cite{mih78}). For the sake of clarity, we reproduced the velocity profile from \cite{hol08} and convolved it with a single Lorentz profile for a gas temperature of $8000$K. In general, when dealing with a non-homogeneous medium, for every position across the planetary disc (see Equation 1), one needs to evaluate the atomic absorption profile locally for the corresponding temperature. The same caution is required when dealing with several populations of particles. As shown in Fig. 7, the velocity distribution field derived by \cite{hol08}, when compared with the same distribution including the effect of natural broadening, leads to distinct results with stronger and extended line wings and a smoothing out of the red side, making the new distribution much symmetric in the star's reference frame  than initially set. Finally, we remark that the effect of natural broadening is even more visible when the ENAs contribution is smaller. \\

With that in mind, one may well imagine a hybrid solution where a thin layer of superthermal atoms (ENAs production, turbulence energized particles, etc.) is on top of a classical thermal atmosphere that may absorb the \lya\ stellar flux in the way it is observed (e.g., Fig. 6). The [H] thickness of the thin layer should not exceed few percent of the column of the underlying atmosphere. This configuration takes its origin in similar scenarios that also occur in the so-called bulge region in the upper atmosphere of Jupiter (\cite{ben93,som95,eme96}).  As previously noted, however, ENAs and ions production as proposed by \cite{hol08} must be evaluated in conjunction with the evolving size of the magnetopause cavity so that the effective ENAs population that contributes to the \lya\ absorption is consistent with the magnetopause position and the gas structure around. Also, the corresponding absorption profile  should account for the effect of natural broadening of the \lya\ line as described here (eg Fig. 7). In practice, many solutions may be found  that may require spanning the stellar and atmospheric models parameters in a self-consistent way. This effort is out of the scope of this study.

\section{ Conclusion} 

HST archive observations of the \lya\ emission of HD209458 are revisited to test the absorption rate sensitivity to the spectral range considered inside the \lya\ emission line and to derive the absorption line profile during transit. Our first conclusion is that the absorption depth during transit is weakly dependent on the assumed spectral range assumed, with rates in the range of $(8.4-8.9)\%\pm 2.\%$ for all possible wavelength windows as reported in our initial study (\bjl). Self-consistent derivation of the absorption line profile during transit shows a symmetrical Lorentzian shape, typical of optically thick conditions of the hydrogen nebula around HD209458b. Whether using direct light curves from the time series or inspecting the absorption line profile during transit, neither method shows any indication of preferred absorption on the blue side of the stellar line in the spectral domain $\sim\pm 200 {\rm km\, s^{-1}}$ from line center.  Finally, we stress that because the planet is not resolved spatially, several scenarios of the gas distribution may lead to the observed absorption line profile. In that context, two scenarios that fit rather well to the absorption profile are suggested. A first model assumes an optically thick atmosphere of HD209458b obtained from the standard model of \cite{mun07}. A second class of models consists of a standard model that may have a thin layer of superthermal neutrals (ENAs, turbulence energized particles, etc.) on top of it, much like what we observe for the upper atmosphere of Jupiter (\cite{ben93,eme96}). In all cases, the natural broadening that was missed in most past studies is required in the modelling of the \lya\ absorption profile. Moreover, for Enas production, a self-consistent treatment of the charge exchange process in conjunction with the size of the magnetospheric cavity is much needed to test the validity of such process. In any case, high resolution observations with a {\it S/N} better than $(0.1-1.0)\% $ are required to efficiently constrain the hydrogen distribution and its time and phase variations along the planetary orbit.

\acknowledgments
 The author acknowledges support from Universit\'e Pierre et Marie Curie (UPMC) and the Centre National de la Recherche Scientifique (CNRS) in France. He also warmly  thanks Prof. W. Harris for his hospitality during his 6 months stay at the Department of Applied Science at UC davis.  This work is based on observations with the NASA/ESA Hubble Space Telescope, obtained at the Space Telescope Science Institute, which is operated by AURA, Inc.


\end{document}